\documentclass[pra,twocolumn,showpacs,superscriptaddress]{revtex4-1}
\usepackage{amsfonts}
\usepackage{amssymb}
\usepackage{amsmath}
\usepackage{graphicx} 
\usepackage{datetime}
\usepackage{enumerate}
\usepackage{multirow}

\usepackage{color}
\usepackage[normalem]{ulem}

\newcommand{\ket}[1]{|#1\rangle}

\begin{document}

\title[Beyond adiabatic elimination]{Beyond adiabatic elimination:\\A hierarchy of approximations for multi-photon processes}
\date{02 April 2013}

\author{Vanessa \surname{Paulisch}}
\email{Vanessa.Paulisch@physik.uni-muenchen.de}
\affiliation{Fakult\"{a}t f\"{u}r Physik, Ludwig-Maximilians-Universit\"{a}t, 
Schellingstrasse 4, 80779 M\"{u}nchen, Germany}

\author{Rui \surname{Han}}
\email{han.rui@quantumlah.org}
\affiliation{Centre for Quantum Technologies, National University of Singapore, 
3 Science Drive 2, Singapore 117543, Singapore} 

\author{Hui Khoon \surname{Ng}}
\email{cqtnhk@nus.edu.sg}
\affiliation{Centre for Quantum Technologies, National University of Singapore, 
3 Science Drive 2, Singapore 117543, Singapore} 
\affiliation{DSO National Laboratories, 20 Science Park Drive, 
Singapore 118230, Singapore}

\author{Berthold-Georg \surname{Englert}}
\email{cqtebg@nus.edu.sg}
\affiliation{Centre for Quantum Technologies, National University of Singapore, 
3 Science Drive 2, Singapore 117543, Singapore} 
\affiliation{Department of Physics, National University of Singapore, 
2 Science Drive 3, Singapore 117542, Singapore}

\begin{abstract}
In multi-level systems, the commonly used adiabatic elimination is a method for
approximating the dynamics of the system by eliminating irrelevant,
non-resonantly coupled levels. 
This procedure is, however, somewhat ambiguous and it is not clear how to
improve on it systematically. 
We use an integro-differential equation for the probability amplitudes of the
levels of interest, which is equivalent to the original Schr\"odinger equation
for all probability amplitudes.
In conjunction with a Markov approximation, the integro-differential equation
is then used to generate a hierarchy of approximations, in which the zeroth order
is the adiabatic-elimination approximation.
It works well with a proper choice of interaction picture; 
the procedure suggests criteria for optimizing this choice.
The first-order approximation in the hierarchy provides significant improvements over standard adiabatic elimination, without much increase in complexity, and is furthermore not so sensitive to the choice of interaction picture. 
We illustrate these points with several examples. 
\end{abstract}


\maketitle

\section{Introduction}

The coherent manipulation of quantum states is central to a large variety of
con\-tem\-po\-rary research in physics (see, for example, Ref.~\cite{CohenTannoudjiBook}), among them experiments
that aim at processing quantum information \cite{NielsenChuang}. 
When implementing a qubit in the internal degrees of freedom of a trapped atom
or ion, the relevant quantum states are different electronic energy levels
manipulated by a combination of static and time-dependent 
electromagnetic fields.  
The ideal case of a two-level system resonantly driven by a laser is rarely
available, as the levels of interest may not be coupled directly by a dipole
transition, or the transition occurs at an inconvenient wavelength.
Instead, two levels are often coupled indirectly through an intermediate level, as in the well-known Raman transition \cite{ScullyBook}. 
More generally, one can couple several levels of interest through multiple intermediate, off-resonant levels, using multiple lasers \cite{Kyrola83, Kyrola87}.
Such a multi-level, multi-photon system is very complicated to analyze exactly, and one usually resorts to numerical solutions.

It is often more insightful, however, to have an analytical, if approximate, solution to the problem. 
For instance, in a three-level Raman process, the intermediate level is far-off-resonant from the laser frequencies, so that it is barely populated during the evolution of the system. 
Other than providing an indirect coupling route to the two states of interest (the \emph{relevant} states), the intermediate state (the \emph{irrelevant} state) hardly affects the dynamics of the relevant states. 
It then seems plausible that one might be able to derive a simpler effective description of the relevant states only, with the irrelevant state eliminated. 
Such is the idea of the commonly used procedure of \emph{adiabatic elimination} (see, for example, Refs.~\cite{ShoreBook,Poizat95,Brion1,Shore08}).

Adiabatic elimination gives an adequate description of the relevant states, as long as the coupling between the relevant and irrelevant states is sufficiently weak, and the number of eliminated states is small, like in a three-level Raman transition.
When eliminating two states in a three-photon transition from a ground state to an excited state through two intermediate far-detuned states, one already sees significant inaccuracies arising from the adiabatic elimination procedure (see the example discussed in Sec.~\ref{sec:EX3photons}). 
In experimental situations, one also often functions in the parameter range where the coupling between relevant and irrelevant states fails to be weak enough for adiabatic elimination to offer good predictions. 
This leaves much to be desired.

In this article, we introduce a systematic hierarchy of approximations for multi-level, multi-photon systems with a clear separation of states into relevant and irrelevant sectors. 
Standard adiabatic elimination emerges as the lowest-order approximation; higher-order approximations give not only more accurate effective descriptions of the relevant states, but also a statement about the evolution of the irrelevant states.
A technical point of consideration is the choice of interaction picture in which conditions of adiabaticity hold (previously discussed in Ref.~\cite{Brion1} for adiabatic elimination in a three-level Raman process); we propose optimality criteria for a judicious choice.
An important conclusion of our studies is that, while one can improve the performance of standard adiabatic elimination with a careful, often complicated, choice of interaction picture, one gains much more in accuracy by proceeding to the next-order approximation in the hierarchy with a near-optimal, but simpler to use, interaction-picture choice.

The article is organized as follows: 
Sec.~\ref{sec:Setup} sets up the problem and introduces the notation. 
We then review the standard method of adiabatic
elimination and the problems that may arise when applying it in Sec.~\ref{sec:AEreview}.
In Sec.~\ref{sec:general}, we present a systematic hierarchy of approximations based on 
integro-differential equations of Lippmann-Schwinger type, and examine the issue of an  appropriate choice of interaction picture. 
In Sec.~\ref{sec:example}, we discuss several examples that illustrate the usefulness of our
approach. 
We close with a summary and outlook in Sec.~\ref{sec:outlook}.

\section{Problem setup: multi-level, multi-photon processes}\label{sec:Setup}

\begin{figure}
\centering
\includegraphics[width=0.45\columnwidth]{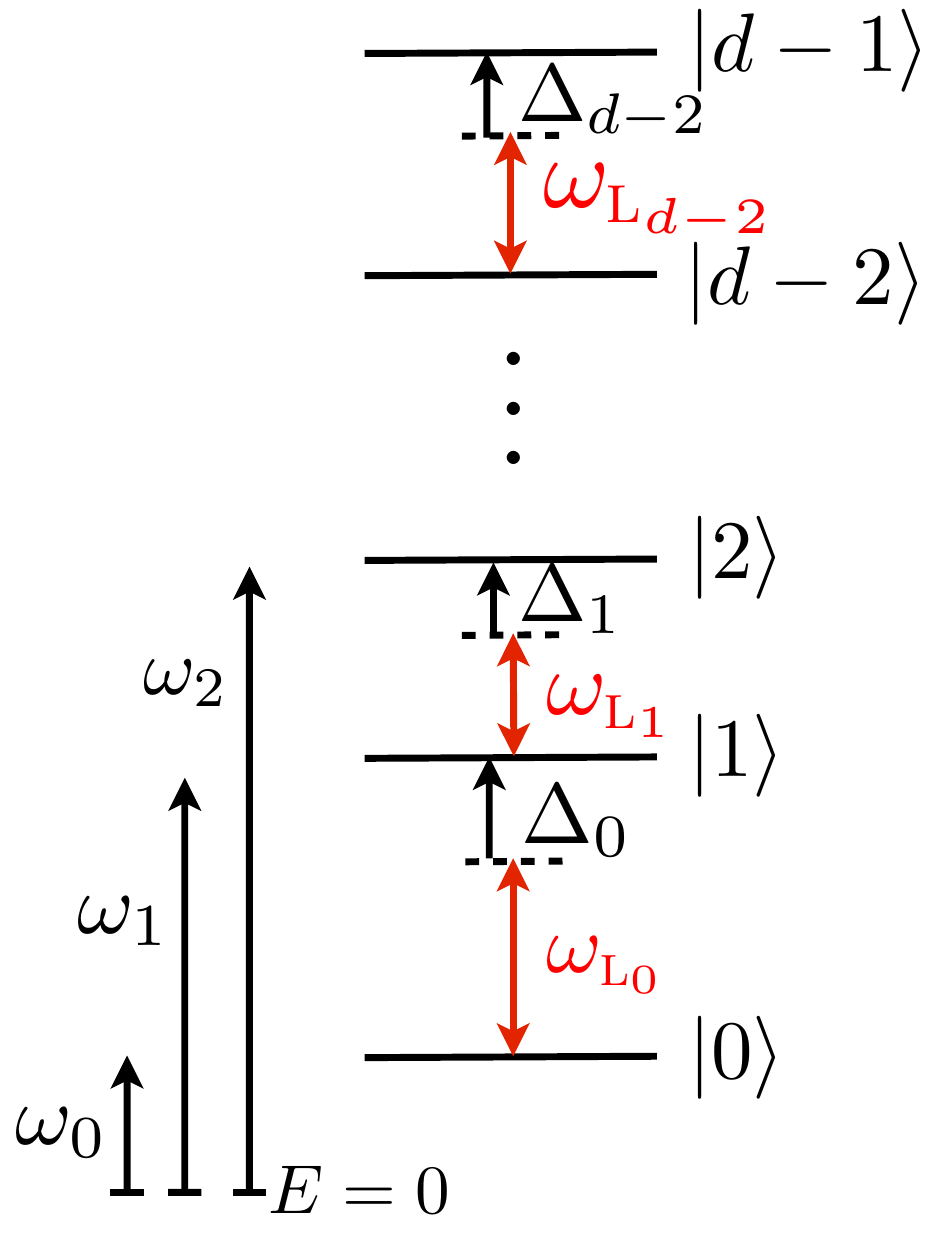}
\caption{
\label{fig:Cascade}
A cascade $d$-level system, with levels pairwise coupled by lasers.}
\end{figure}

Consider a system of $d$ levels, $\{|0\rangle,|1\rangle,\ldots,|d-1\rangle\}$, numbered such that (dipole) transition between each pair of states $|i\rangle$ and $|i+1\rangle$, for $i=0,\ldots, d-2$, is driven by a laser with frequency $\omega_{\textsc{l}_i}(>0)$. 
An example is the cascade system illustrated in Fig.~\ref{fig:Cascade}.
For such a $d$-level system, the Hamiltonian (under the rotating-wave approximation) is
\begin{equation}\label{eq:Hnlevel}
H=\hbar\sum_{i=0}^{d-1}|i\rangle\,\omega_i\,\langle i|+\frac{\hbar}{2}\sum_{i=0}^{d-2}\Omega_i{\Big(|i\rangle\,\mathrm{e}^{\mathrm{i}q_i\omega_{\textsc{l}_i}t}\,\langle i+1|+\textrm{h.c.}\Big)}
\end{equation}
Here, $\hbar\omega_i$ is the energy (with respect to some reference) of $|i\rangle$, and $q_i$ is a binary parameter: $q_i=1$ if $\omega_{i+1}>\omega_i$ (absorb a photon to go from $|i\rangle$ to $|i+1\rangle$); $q_i=-1$ if $\omega_{i+1}<\omega_i$ (emit a photon to go from $|i\rangle$ to $|i+1\rangle$). 
$\Omega_i(>0)$ 
\footnote{For simplicity of notation, we assume $\Omega_i>0$, but this is inconsequential to our conclusions, and one can easily restore the appropriate phases of the $\Omega_i$s if needed. 
In all our examples, $\Omega_i$ can always be chosen as positive by absorbing any phase arising from the dipole-matrix element in the definition of the basis states.} 
is the Rabi frequency for transition between $|i\rangle$ and $|i+1\rangle$, which depends on the amplitude and the polarization of the
electric field, as well as the dipole matrix element between the states. 

It is simplest to work in an interaction picture where the Hamiltonian is time independent, if such a picture exists.
This is always possible for the $d$-level system described above, by going to the interaction picture defined by a Hamiltonian
\begin{equation}\label{eq:H0Nlevel}
H_0=|0\rangle\,\hbar\omega_0\,\langle 0|+\sum_{i=1}^{d-1}|i\rangle{\left(\hbar\omega_0+\sum_{k=0}^{i-1}q_k \hbar\omega_{\textsc{l}_k}\right)}\langle i|\,.
\end{equation}
The interaction-picture Hamiltonian is then
\begin{eqnarray}\label{eq:HINlevel} 
H_\mathrm{I}&=&\mathrm{e}^{\mathrm{i}H_0t/\hbar}(H-H_0)\mathrm{e}^{-\mathrm{i}H_0t/\hbar}\nonumber\\
&=&\hbar{\left(
\begin{array}{@{}ccccc@{}}
0&\frac{1}{2}\Omega_{0}&0&\ldots&0\\
\frac{1}{2}\Omega_{0}&q_0\Delta_0&\frac{1}{2}\Omega_1&\ldots&0\\
0&\frac{1}{2}\Omega_{1}&q_0\Delta_0+q_1\Delta_1&\ldots&0\\
\vdots&\vdots&\vdots&\ddots&\vdots\\
0&0&0&\ldots&~\displaystyle{\sum_{i=0}^{d-2}q_i\Delta_i}
\end{array}
\right)},~
\end{eqnarray}
written as a matrix in the basis $\{|i\rangle\}_{i=0}^{d-1}$.
Here, $\Delta_i$ is the laser detuning for the transition between states $|i\rangle$ and $|i+1\rangle$,
\begin{equation}
\Delta_i=|\omega_{i+1}-\omega_i| - \omega_{\textsc{l}_i}=q_i(\omega_{i+1}-\omega_i) -\omega_{\textsc{l}_i}.
\end{equation}

A different choice of $H_0$ gives a different interaction picture; a generic choice gives $H_\mathrm{I}$ with an explicit time dependence. 
Since it is simpler to work with a time-independent $H_\mathrm{I}$, we restrict our attention only to such cases. 
However, even the choice of time-independent interaction picture is nonunique.
Starting from any time-independent interaction picture, like the one determined by $H_0$ of Eq.~\eqref{eq:H0Nlevel}, it is always possible to go to another time-independent interaction picture by
subtracting a multiple of the identity from $H_0$, i.e., $H_0\rightarrow H_0-\hbar \widetilde\omega$ for any frequency $\widetilde\omega$.
This yields a different interaction-picture Hamiltonian, 
$H_\mathrm{I} \rightarrow H_\mathrm{I} + \hbar\widetilde\omega$. 
For exact solutions of the problem, all interaction pictures are equivalent; for approximate solutions, as we shall see below, the choice of interaction picture in which the approximations are made affects the accuracy of the solution.

For the system in question, we suppose that there is a clear separation of the states into a \emph{relevant} sector and an \emph{irrelevant} sector.
The relevant sector must contain at least the initial state of the system, and all states that are near-resonantly or strongly connected to the initial state; the irrelevant sector contains only states that play a small role in the dynamics of the system because they are far-off-resonantly and weakly coupled to any of the states in the relevant sector.
More precisely, we assume that we can rearrange $H_\mathrm{I}$, written in some appropriate interaction picture (the choice of which will be made clear in Sec.~\ref{sec:general}), into submatrices $\boldsymbol{\omega},\boldsymbol{\Omega}$ and $\boldsymbol{\Delta}$,
\begin{equation} \label{eq:HintPsi}
 H_\mathrm{I} = \hbar{\left(
 \begin{array}{@{}cc@{}} \boldsymbol{\omega} & \frac{1}{2}\boldsymbol{\Omega} \\ 
 \frac{1}{2}\boldsymbol{\Omega}^\dagger &\boldsymbol{\Delta} 
\end{array} 
\right)},
 \end{equation}
where $\boldsymbol{\omega}$ involves only the relevant states (say $m$ of them), and $\boldsymbol{\Delta}$ involves only the irrelevant states (say $n=d-m$ of them). 
$\boldsymbol{\Omega}$ connects the relevant and irrelevant sectors and is assumed to be small compared with $\boldsymbol{\Delta}$.

The probability amplitudes of the (interaction-picture) states are collected into a $m$-component column $\psi(t)$ for the relevant states, and a $n$-component column $\epsilon(t)$ for the irrelevant states so that 
\begin{equation}\label{eq:Psi}
\Psi(t)={\left(
\begin{array}{@{}c@{}} \psi(t) \\ \epsilon(t)\end{array}
\right)}
\end{equation}
is the $d$-component column of probability amplitudes.
The Schr\"{o}dinger equation then takes the form
\begin{subequations}\label{eq:SEcomp}
\begin{eqnarray}
\mathrm{i} \frac{\partial}{\partial t} \psi (t)&=& \boldsymbol{\omega} \psi (t) + \frac{1}{2}\boldsymbol{\Omega} \epsilon (t)\label{eq:SEcompa}\\
\mathrm{i} \frac{\partial}{\partial t}  \epsilon (t)&=&\frac{1}{2} \boldsymbol{\Omega}^\dagger \psi (t)+ \boldsymbol{\Delta}\epsilon (t).\label{eq:SEcompb}
\end{eqnarray}
\end{subequations}
In the spirit of adiabatic elimination, we seek a good
effective description for the relevant amplitudes $\psi(t)$, while being content with a less accurate approximation for the irrelevant amplitudes $\epsilon(t)$.

For the sake of concreteness, we have described the formalism so far in terms of a system with $d$ states, pairwise coupled by lasers.
However, our discussion below applies to any situation in which one can arrive at an interaction Hamiltonian that is time-independent, and for which there is a clear separation into relevant and irrelevant sectors weakly coupled to each other.
For situations where the time dependence of the dynamics cannot be transformed away by an suitable choice of interaction picture (for example, in the case of time-dependent Rabi frequencies), we refer the reader to Refs.~\cite{Torosov09,Torosov12}.

\section{Adiabatic elimination}\label{sec:AEreview}

\subsection{A brief review}
\begin{figure}
\centering
\includegraphics[width=0.57\columnwidth]{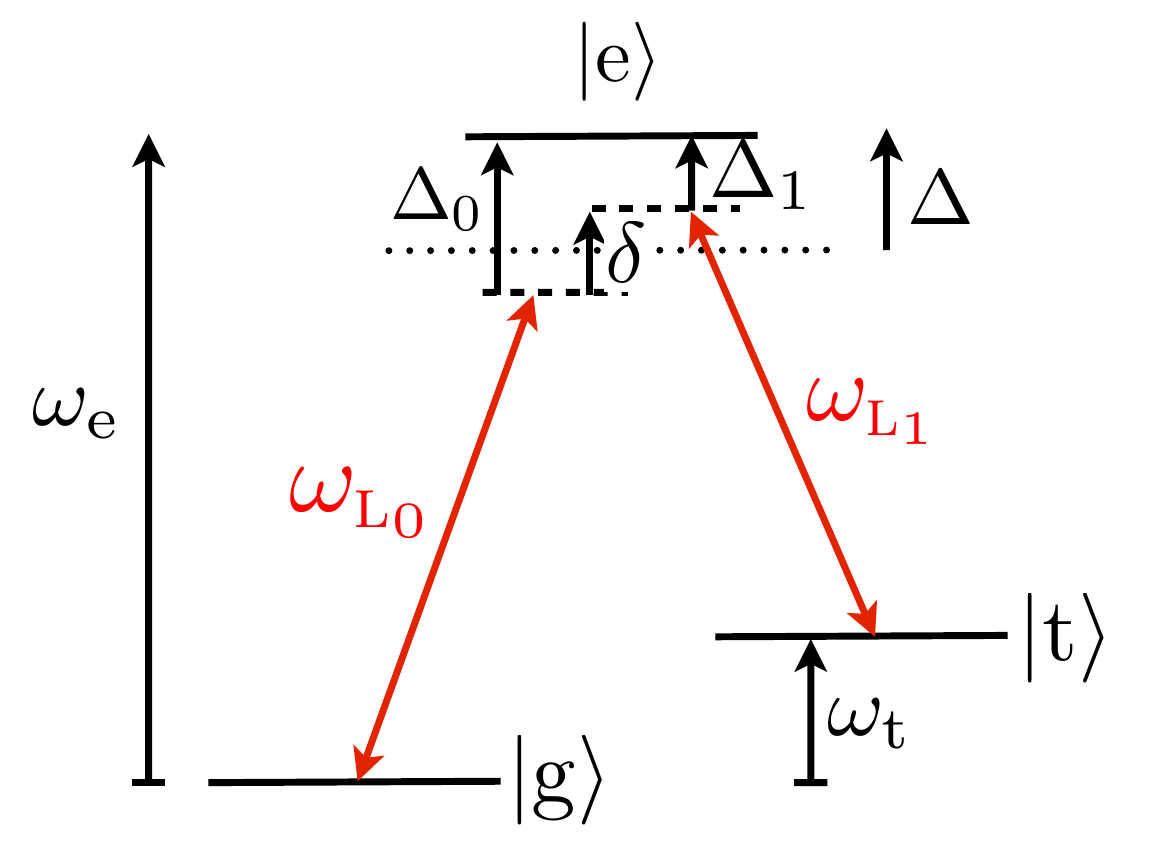}
\caption{
\label{fig:Lambda}
The Raman process in a three-level system couples an initial state $|\mathrm{g}\rangle$ to a target state $|\mathrm{t}\rangle$ 
via a far-detuned intermediate state $|\mathrm{e}\rangle$, with respective energies $\hbar\omega_{\mathrm{g}},\hbar\omega_{\mathrm{t}}$, and $\hbar\omega_\mathrm{e}$.
The system is in a $\Lambda$ configuration ($\omega_{\mathrm{g}},\omega_{\mathrm{t}}<\omega_\mathrm{e}$, illustrated here), a V configuration ($\omega_{\mathrm{g}},\omega_{\mathrm{t}}>\omega_\mathrm{e}$), or a cascade configuration ($\omega_{\mathrm{g}}<\omega_\mathrm{e}<\omega_{\mathrm{t}}$).
In all three configurations, the numbering convention of Sec.~\ref{sec:Setup} assigns $|0\rangle\equiv |\mathrm{g}\rangle$, $|1\rangle\equiv |\mathrm{e}\rangle$ and $|2\rangle\equiv |\mathrm{t}\rangle$.
The transitions are driven by two lasers of frequencies $\omega_{\mathrm{L_0}}$ and
$\omega_{\mathrm{L_1}}$, with respective detunings (for the $\Lambda$ configuration) \mbox{$\Delta_0=\omega_\mathrm{e}-\omega_{\textsc{l}_0}$} and \mbox{$\Delta_1=(\omega_{\mathrm{e}}-\omega_{\mathrm{t}})-\omega_{\textsc{l}_1}$}.
$\Delta_0$ and $\Delta_1$ are large in magnitude by design, to avoid populating the $|\mathrm{e}\rangle$ state; their difference is small by design, so that the two-photon transition from $|\mathrm{g}\rangle$ to $|\mathrm{t}\rangle$ is nearly resonant. 
}
\end{figure}

Adiabatic elimination is usually discussed in the context of a three-level Raman process, like that illustrated in Fig.~\ref{fig:Lambda}, to eliminate the irrelevant intermediate state $|\mathrm{e}\rangle$ which provides the indirect coupling between the two states of interest $|\mathrm{g}\rangle$ and $|\mathrm{t}\rangle$.
The time evolution of such a three-level problem can, of course, be studied using an exact eigen-decomposition, but the resulting expressions for the eigenvalues and eigenvectors of $H_{\mathrm{I}}$ are often involved and not transparent. 
Instead, adiabatic elimination is used to remove the intermediate state from the description, which is possible whenever $|\mathrm{e}\rangle$ is only weakly and far-off-resonantly coupled to $|\mathrm{g}\rangle$ and $|\mathrm{t}\rangle$, i.e., $|\mathrm{e}\rangle$ is an irrelevant state.
The three-level system is reduced to a two-level system of relevant states, with a $2\times2$ matrix for its effective Hamiltonian.
Then, all the familiar tools for two-level systems are applicable, and one can easily work out the frequency and amplitude of the oscillatory probability amplitudes as well as other details of interest. 

More generally, one can perform adiabatic elimination on a $d$-level system to take care of the irrelevant sector and reduce  the description to an effective Hamiltonian for the relevant states only.
The adiabatic prescription amounts to setting the time derivative of the irrelevant state amplitude to zero, that is,
\begin{equation}
\mathrm{i} \frac{\partial}{\partial t} \epsilon (t) = 
\frac{1}{2} \boldsymbol{\Omega}^\dagger\psi(t)
+ \boldsymbol{\Delta} \epsilon(t) 
\approx 0,
\end{equation}
from Eq.~\eqref{eq:SEcompb} for the interaction-picture Hamiltonian of Eq.~\eqref{eq:HintPsi}.
Assuming that $\boldsymbol{\Delta}$ is invertible (as is usually true in cases of physical interest), this gives
\begin{equation}\label{eq:AEep}
\epsilon(t)\approx -\frac{1}{2\boldsymbol{\Delta}}\boldsymbol{\Omega}^\dagger\psi(t).
\end{equation}
After using this to eliminate $\epsilon(t)$ from the equation of motion for $\psi(t)$ [Eq.~\eqref{eq:SEcompa}], we obtain
\begin{equation}
\mathrm{i}\frac{\partial}{\partial t}\psi(t)={\left(\!\boldsymbol{\omega}-\boldsymbol{\Omega}\frac{1}{4\boldsymbol{\Delta}}\boldsymbol{\Omega}^\dagger\!\right)}\psi(t),
\end{equation}
and the effective Hamiltonian for the relevant sector can be read off as
\begin{equation}\label{eq:HeffNlevel}
H_\mathrm{eff}=\hbar{\left(\!\boldsymbol{\omega}-\boldsymbol{\Omega}\frac{1}{4\boldsymbol{\Delta}}\boldsymbol{\Omega}^\dagger\!\right)}.
\end{equation}

As an example, we recall the adiabatic elimination procedure for the $\Lambda$-configuration of Fig.~\ref{fig:Lambda}. 
This is usually done in a slightly different interaction picture than the one determined by $H_0$ of Eq.~\eqref{eq:H0Nlevel}. Setting $\widetilde\omega=\frac{1}{2}\delta$,  the Hamiltonian $H_\mathrm{I}$ in the interaction picture for $H_0-\frac{1}{2}\delta$ is
\begin{equation}\label{eq:HI}
H_\mathrm{I}=\hbar{\left(
\begin{array}{@{}cc|c@{}}
-\frac{1}{2}\delta & 0 &  \frac{1}{2}\Omega_0 \\[0.5ex]
0 &  \frac{1}{2}\delta &  \frac{1}{2}\Omega_1 \\[0.5ex]
\hline
\frac{1}{2}\Omega_0 & \frac{1}{2}\Omega_1 & \Delta
\end{array}
\right)}
= \hbar{\left(
\begin{array}{@{}c|c@{}} 
\boldsymbol{\omega} & \frac{1}{2}\boldsymbol{\Omega} \\[0.5ex]
\hline 
\frac{1}{2}\boldsymbol{\Omega}^\dagger & \boldsymbol{\Delta}
\end{array} 
\right)},
\end{equation}
written in the basis $\{|\mathrm{g}\rangle,|\mathrm{t}\rangle, |\mathrm{e}\rangle\}$.
Here, $\delta$ is the overall detuning of the two-photon Raman process,
\begin{equation}
\delta=\Delta_0-\Delta_1 =\omega_{\mathrm{t}}-(\omega_{\textsc{l}_0}-\omega_{\textsc{l}_1}),
\end{equation}
and $\Delta$ is the average detuning of the intermediate state,
\begin{equation}
\Delta=\frac{1}{2}(\Delta_0+\Delta_1).
\end{equation}
For a typical Raman process, we have $|\delta|\ll|\Delta|$, and $\Omega_0,\Omega_1\ll|\Delta|$ so that $|\mathrm{e}\rangle$ is weakly coupled to $|\mathrm{g}\rangle$ and $|\mathrm{t}\rangle$.
Adiabatic elimination of $|\mathrm{e}\rangle$ then gives an effective Hamiltonian
\begin{equation}\label{eq:Heff}
H_\mathrm{eff} = -\frac{\hbar}{2}{\left(
\begin{array}{@{}cc@{}} 
\delta + \frac{ | \Omega_0 | ^2 }{ 2 \Delta} & \frac{ \Omega_0 \Omega_{1} }{2 \Delta} \\[1ex]
\frac{ \Omega_1\Omega_0  }{ 2 \Delta } & -\delta + \frac{ | \Omega_1 | ^2 }{ 2 \Delta }
\end{array} 
\right)},
\end{equation}
now written in the basis $\{|\mathrm{g}\rangle,|\mathrm{t}\rangle\}$ that involves the relevant states only. 
This effective Hamiltonian describes simple Rabi oscillations between the two relevant states, and one can obtain the effective Rabi frequency and the population of the relevant states as a function of time.

\subsection{Problems with adiabatic elimination}\label{sec:problems}
Despite its popular use in simplifying multi-level problems, the procedure of adiabatic elimination has certain
ambiguities and potential problems. 
We present these in the form of four questions:
\begin{enumerate}[(a)]
\item \label{Qa}
  The basic assumption that the time derivative of $\epsilon(t)$
  is small and can be treated as approximately vanishing is questionable.
  Even if the population of the irrelevant states is small, its time
  derivative need not be small; in fact, it can be rather sizeable
  because the large detuning makes $\epsilon(t)$ oscillate rapidly. 
  \emph{Is there a better justification for this assumption?}
\item \label{Qc}
  There are problems with the normalization of the wave function.
  Originally, we have $\psi(t)^\dagger\psi(t)+\epsilon(t)^\dagger\epsilon(t)=1$.
  As $H_\mathrm{eff}$ is hermitian, evolution under $H_\mathrm{eff}$ preserves the normalization of $\psi(t)$, i.e., $\psi(t)^\dagger\psi(t)=\textrm{constant}$ for all $t$.
  Combined with the initial condition that all population is contained in $\psi(t=0)$,  
  this implies $\epsilon(t)= 0$ for all $t$, whereas the basic
  approximation in (\ref{eq:AEep}) says otherwise.
  Thus adiabatic elimination fails to make a consistent statement about the population in
  the intermediate states. 
  \emph{Is it possible to estimate the population in the eliminated states \emph{and}
  comply with the normalization conditions?} 
\item \label{Qb}
  As pointed out in Sec.~\ref{sec:Setup}, the choice of interaction picture is nonunique.
  Applying the adiabatic elimination procedure in a different interaction picture yields a different effective
  Hamiltonian and thus different predictions for the evolution of the relevant states.
  There have been attempts to identify the ``correct'' choice of interaction
  picture, which gave valuable insights~\cite{Brion1}, but a
  definite answer is still lacking.
  \emph{How can one choose an optimal interaction picture?}
\item \label{Qd}
  For larger Rabi frequencies or smaller detunings, or a system with several irrelevant states, one observes that the
  adiabatic elimination does not give a trustworthy approximation.
  \emph{Is there a systematic way of improving the accuracy of the adiabatic-elimination approximation?}
\end{enumerate}
We offer answers to these four questions in the next section.

\section{A systematic hierarchy of approximations}\label{sec:general}

We begin with the equations of motion \eqref{eq:SEcomp}, with the initial condition $\epsilon(t=0)=0$.
The differential equation \eqref{eq:SEcompb} for $\epsilon(t)$ can be solved exactly,
\begin{equation}
\epsilon (t) = - \frac{\mathrm{i}}{2} \int\limits_0^t \mathrm{d} t'\, 
\mathrm{e}^{-\mathrm{i} \boldsymbol{\Delta} (t-t')} \boldsymbol{\Omega}^\dagger \psi(t').
\label{eq:epsexact}
\end{equation}
We then use this in the differential equation \eqref{eq:SEcompa} for $\psi(t)$ to arrive at an
integro-differential equation of Lippmann-Schwinger type,
\begin{equation}
  \mathrm{i}  \frac{\partial}{\partial t} \psi(t) = 
  \boldsymbol{\omega} \psi (t) - \frac{\mathrm{i}}{4} \boldsymbol{\Omega} \int\limits_0^t \mathrm{d} t'\, 
   \mathrm{e}^{-\mathrm{i} \boldsymbol{\Delta} (t-t')} \boldsymbol{\Omega}^\dagger \psi(t').
\label{eq:psiexact}
\end{equation}
Together with Eq.~\eqref{eq:epsexact}, this is fully equivalent to Eqs.~\eqref{eq:SEcomp}.
While no approximation entered in the transition from Eqs.~\eqref{eq:SEcomp} to
Eq.~\eqref{eq:psiexact}, the integro-differential equation now serves as the
starting point for the generation of a hierarchy of approximations to solve for $\psi(t)$, with the adiabatic-elimination answer as the zeroth-order solution.

\subsection{Zeroth-order Markov approximation}
The usual (zeroth-order) Markov approximation assumes that memory effects are
negligible, so that $\psi(t') \approx \psi (t)$ in the integral in Eq.~\eqref{eq:psiexact}. 
This assumption is valid, provided that $\psi (t)$ oscillates much more slowly
than $\mathrm{e}^{-\mathrm{i} \boldsymbol{\Delta} t}$.
This happens whenever $\boldsymbol{\Omega}$ and $\boldsymbol{\omega}$, which govern the evolution of $\psi(t)$, contain frequencies that are small on the scale set by $\boldsymbol{\Delta}$. 
We can write this formally as $\Vert\boldsymbol{\Omega}\Vert, \Vert\boldsymbol{\omega}\Vert \ll \Vert\boldsymbol{\Delta}\Vert$, where $\Vert\cdot\Vert$ denotes any norm for operators.
The condition $\Vert\boldsymbol{\omega}\Vert\ll\Vert\boldsymbol{\Delta}\Vert$ requires the appropriate choice of interaction picture (we will return to this point momentarily), while $\Vert\boldsymbol{\Omega}\Vert\ll\Vert\Delta\Vert$ is the statement that the
irrelevant states are weakly and non-resonantly coupled to the relevant states. 
As discussed in standard texts (for example, Refs.~\cite{Haken,CohenT}), a coarse graining is then appropriate in the evaluation of the remaining integral over $t'$  because time scales of order $\Vert\boldsymbol{\Delta}\Vert^{-1}$ are
not resolved and $\mathrm{e}^{- \mathrm{i} \boldsymbol{\Delta} t}$ averages out over the coarse-grained time intervals.
Accordingly, we have 
\begin{equation}\label{eq:intapprox}
\int\limits_0^t \mathrm{d} t' \,\mathrm{e}^{-\mathrm{i} \boldsymbol{\Delta} (t-t')} = 
\frac{1-\mathrm{e}^{-\mathrm{i} \boldsymbol{\Delta} t}}{\mathrm{i} \boldsymbol{\Delta}} 
\approx \frac{1}{\mathrm{i} \boldsymbol{\Delta}},
\end{equation}
and thus obtain the effective Hamiltonian for the zeroth-order Markov approximation,
\begin{equation}\label{eq:0thHeff}
H_\mathrm{eff}^\mathrm{(0)} 
=\hbar{\left(\boldsymbol{\omega} - \boldsymbol{\Omega} \frac{1}{4\boldsymbol{\Delta}} \boldsymbol{\Omega}^\dagger \right)}.
\end{equation}
This is exactly the effective Hamiltonian that adiabatic
elimination gave us [Eq.~\eqref{eq:HeffNlevel}]. 
In this sense then, we have given an answer to question (\ref{Qa}) in Sec.~\ref{sec:problems}: 
Setting $\mbox{\small$\displaystyle\frac{\partial}{\partial t}$}$$\epsilon(t)\approx 0$ in the adiabatic elimination procedure
amounts to a shorthand for the above coarse graining in time to discard rapidly oscillating features
and retain only the slowly varying dynamics.
This justification of the adiabatic elimination procedure as a zeroth-order Markov approximation is also discussed in Ref.~\cite{ShoreBook}.

\subsection{First-order Markov approximation and beyond}
To gain some information about the evolution of $\epsilon(t)$, and so arrive
at an answer to question (\ref{Qc}), we consider the first-order Markov
approximation that takes a bit of the history of the relevant states into
account in the integral in Eq.~\eqref{eq:psiexact}. 
We accomplish this by means of the first-order approximation
\begin{equation}\label{eq:psi1st}
\psi(t') \approx \psi(t) - (t-t')\frac{\partial\psi (t)}{\partial t}
\end{equation}
from a Taylor-series expansion about $t'=t$.
With Eq.~\eqref{eq:intapprox} and 
\begin{equation}
\int\limits_0^t \mathrm{d} t'\, (t-t') \mathrm{e}^{-\mathrm{i} \boldsymbol{\Delta} (t-t')}
\approx -\frac{1}{\boldsymbol{\Delta}^2},
\end{equation}
this gives
\begin{equation}
\mathrm{i} \frac{\partial\psi (t)}{\partial t} = 
{\left(\!\boldsymbol{\omega} - \boldsymbol{\Omega} \frac{1}{4\boldsymbol{\Delta}} \boldsymbol{\Omega}^\dagger \!\right)} \psi(t) 
- \mathrm{i} \boldsymbol{\Omega} \frac{1}{4\boldsymbol{\Delta}^2} \boldsymbol{\Omega}^\dagger
\frac{\partial\psi(t)}{\partial t}.
\end{equation}
When defining the effective Hamiltonian, we now have a choice between
\begin{eqnarray}
\mathrm{i} \frac{\partial\psi (t)}{\partial t} &=&\frac{1}{\hbar} H_{\mathrm{eff}}^{\mathrm{(1)}}\psi(t)\label{eq:2ndM-1}
\end{eqnarray}
and
\begin{eqnarray}
\mathrm{i} \frac{\partial\psi'(t)}{\partial t}&=&\frac{1}{\hbar} H_{\mathrm{eff}}^{\mathrm{(1')}}\psi'(t)\,,\label{eq:2ndM-2}
\end{eqnarray}
where
\begin{equation}\label{eq:2ndM-1a}
\frac{H_{\mathrm{eff}}^{\mathrm{(1)}}}{\hbar}={\left(\!1 + \boldsymbol{\Omega} \frac{1}{4\boldsymbol{\Delta}^2} \boldsymbol{\Omega}^\dagger\!\right)}^{-1}
{\left(\! \boldsymbol{\omega} - \boldsymbol{\Omega} \frac{1}{4\boldsymbol{\Delta}} \boldsymbol{\Omega}^\dagger \!\right)},\end{equation}
and
\begin{align}
\frac{H_{\mathrm{eff}}^{\mathrm{(1')}}}{\hbar} \!&=\!\!
{\left(\! 1 \!+ \!\boldsymbol{\Omega} \frac{1}{4\boldsymbol{\Delta}^{\!2}} \boldsymbol{\Omega}^\dagger \!\right)}^{\!\!\!-\!\frac{1}{2}}
\!{\left(\! \boldsymbol{\omega}\!-\!\boldsymbol{\Omega} \frac{1}{4\boldsymbol{\Delta}} \boldsymbol{\Omega}^\dagger \!\right)}
 {\left(\!1 \!+ \!\boldsymbol{\Omega} \frac{1}{4\boldsymbol{\Delta}^{\!2}} \boldsymbol{\Omega}^\dagger \!\right)}^{\!\!\!-\!\frac{1}{2}}\!\!\!,\nonumber\\[0.5ex]
\psi'(t)
&=\!\!\left(\!1+\boldsymbol{\Omega}\frac{1}{4\boldsymbol{\Delta}^2}\boldsymbol{\Omega}^\dagger\!\right)^{\!\!-\frac{1}{2}}\psi(t).  \label{eq:2ndM-2a}
\end{align}
The effective Hamiltonian in Eqs.~\eqref{eq:2ndM-2a} describes the evolution of the dressed wave function $\psi'(t)$, and is hermitian in the usual sense, so that
\begin{equation}\label{eq:2ndM-2c}
\psi'(t)^{\dagger}\psi'(t)
=\psi(t)^{\dagger}{\left(\!1+\boldsymbol{\Omega}\frac{1}{4\boldsymbol{\Delta}^2}\boldsymbol{\Omega}^\dagger\!\right)}\psi(t)
\end{equation}
is constant in time.
By contrast, the effective Hamiltonian in Eq.~\eqref{eq:2ndM-1a} is hermitian for
the modified inner product 
\begin{equation}\label{eq:2ndM-1'}
(\psi_1 , \psi_2 )\equiv 
\psi_1^\dagger {\left(\!1+\boldsymbol{\Omega}\frac{1}{4\boldsymbol{\Delta}^2}\boldsymbol{\Omega}^\dagger\!\right)}\psi_2,
\end{equation}
giving the right-hand side of Eq.~\eqref{eq:2ndM-2c} for the normalization of
$\psi(t)$ in accordance with ${(\psi,\psi)=1}$.
It does not matter which formulation we prefer for the first-order
Markov approximation, either Eq.~\eqref{eq:2ndM-1} with Eqs.~\eqref{eq:2ndM-1a} and \eqref{eq:2ndM-1'}, or
Eq.~\eqref{eq:2ndM-2} with Eqs.~\eqref{eq:2ndM-2a} is fine.
The two formulations are related to each other by the similarity
transformation afforded by $(1+\frac{1}{4}\boldsymbol{\Omega}\boldsymbol{\Delta}^{-2}\boldsymbol{\Omega}^{\dagger})^{1/2}$, and reliable results can only be expected if this operator does not differ much from the identity.
This gives a precise meaning to the relevant/irrelevant-sector separation requirement that
the coupling part $\boldsymbol{\Omega}$ of the interaction-picture Hamiltonian in
Eq.~\eqref{eq:HintPsi} should be small on the scale set by $\boldsymbol{\Delta}$.

A comparison of Eq.~\eqref{eq:2ndM-2c} with the normalization of the full column
$\Psi$ of Eq.~\eqref{eq:Psi}, 
\begin{equation}\label{eq:PsiPsi}
 \Psi^\dagger \Psi = \psi^\dagger \psi + \epsilon^\dagger \epsilon =  1,
\end{equation}
reveals that $\epsilon^{\dagger}\epsilon$ is here approximated by
\begin{equation}\label{eq:epseps}
\epsilon^{\dagger} \epsilon
\approx\psi^{\dagger}\boldsymbol{\Omega}\frac{1}{4\boldsymbol{\Delta}^2}\boldsymbol{\Omega}^{\dagger}\psi,
\end{equation}
which is just what the zeroth-order Markov approximation says,
\begin{equation}
\epsilon (t) = -\frac{\mathrm{i}}{2} \int\limits_0^t \mathrm{d} t'\, 
\mathrm{e}^{-\mathrm{i} \boldsymbol{\Delta} (t-t')}\boldsymbol{\Omega}^{\dagger}\psi(t)
\approx - \frac{1}{2\boldsymbol{\Delta}} \boldsymbol{\Omega}^{\dagger} \psi(t).
\end{equation}
This observation is reassuring and provides the answer to question (\ref{Qc})
in Sec.~\ref{sec:problems}.

The first-order Markov approximation offers a correction to adiabatic elimination.
By including another term in the Taylor-series approximation
in Eq.~(\ref{eq:psi1st}), one gets a second-order Markov approximation. 
A second time derivative of $\psi(t)$ will appear, but it can be
approximated by replacing one of the time derivatives by an application of the
effective Hamiltonian, thereby obtaining a Schr\"odinger-type first-order
differential equation for $\psi(t)$. 
Successive terms in the Taylor series give a systematic hierarchy of approximations, with adiabatic elimination as the zeroth-order  procedure, and this gives an answer to question (\ref{Qd}).
As we shall see shortly, however, the first-order Markov approximation is
already of sufficient quality, so that the complications of higher-order
approximations are hardly worth the trouble.

\subsection{Choice of interaction picture}
There still remains question (\ref{Qb}) about the choice of a good interaction picture that leads to an effective Hamiltonian with the correct physics.
The effective Hamiltonians 
$H_{\mathrm{eff}}^{\mathrm{(0)}}$ and
$H_{\mathrm{eff}}^{\mathrm{(1)}}$  of Eqs.~\eqref{eq:0thHeff} and \eqref{eq:2ndM-1a} take the form
\begin{eqnarray}
\widetilde{H}_{\mathrm{eff}}^{\mathrm{(0)}} 
&=&\hbar{\left(\!\boldsymbol{\omega} + \widetilde{\omega} 
- \boldsymbol{\Omega} \frac{1}{4(\boldsymbol{\Delta} + \widetilde{\omega})} \boldsymbol{\Omega}^{\dagger}\!\right)}, \nonumber\\[1ex]
\widetilde{H}_{\mathrm{eff}}^{\mathrm{(1)}} 
&=& {\left(\!1 + \boldsymbol{\Omega} \frac{1}{4(\boldsymbol{\Delta} + \widetilde{\omega})^2} 
\boldsymbol{\Omega}^{\dagger}\!\right)}^{-1}\,
\widetilde{H}_{\mathrm{eff}}^{\mathrm{(0)}}
\label{eq:HeffM}
\end{eqnarray}
in an alternate interaction picture specified by a shift of $\hbar\widetilde{\omega}$.

We recall that the crucial assumption in the Markov approximation is that
$\psi(t)$ oscillates slowly in comparison with 
$\mathrm{e}^{-\mathrm{i}\boldsymbol{\Delta} t}$, which is invoked when replacing $\psi(t')$
by $\psi(t)$ on the right-hand side of (\ref{eq:psi1st}) in the
integro-differential equation (\ref{eq:psiexact}).  
Consistency,  therefore, requires that the magnitudes of the eigenvalues of the effective
Hamiltonian should be as small as possible because they determine
the frequencies contained in $\psi(t)$.
The following three different conditions for choosing $\widetilde{\omega}$ suggest themselves: 
\begin{subequations}\label{eq:optcond}
\begin{eqnarray}
&&\mathrm{tr}\bigl\{\boldsymbol{\omega} + \widetilde{\omega}\bigr\}=0,\label{conda}\\
&&\Vert\widetilde{H}_{\mathrm{eff}}\Vert_\mathrm{op} \mbox{ is minimal},\label{condb}\\
\textrm{or}\quad&&\Vert\widetilde{H}_{\mathrm{eff}}\Vert_\mathrm{tr}=\mathrm{tr}\{\vert\widetilde{H}_{\mathrm{eff}}\vert\}\mbox{ is minimal}\label{condc}.
\end{eqnarray}
\end{subequations}
Here, $\Vert H\Vert_\mathrm{op}$ refers to the operator norm of $H$, while $\Vert H\Vert_\mathrm{tr}$ is the trace norm \footnote{For an operator $O$ with eigenvalues $\lambda_i$, $|| O||_{\mathrm{op}}={\max}_i |\lambda_i|$, and $|| O ||_{\mathrm{tr}}=\sum_i |\lambda_i|$.}.
Conditions \eqref{condb} and \eqref{condc} insist on choices of $\widetilde{\omega}$ such that $\widetilde{H}_\mathrm{eff}$ has eigenvalues that are small in two different senses:
The operator-norm condition requires minimizing the largest absolute value of the eigenvalues of $\widetilde H_\mathrm{eff}$; the trace-norm condition requires minimizing the sum of the absolute values of the eigenvalues of $\widetilde H_\mathrm{eff}$.
Condition \eqref{conda} prescribes distributing the eigenvalues of the $\boldsymbol{\omega}$ part of the interaction-picture Hamiltonian of Eq.~\eqref{eq:HintPsi} about 0.
This does not necessarily guarantee small eigenvalues,
but for the Hamiltonians in Eqs.~\eqref{eq:HeffM}, a different choice of $\widetilde{\omega}$ gives, as the dominant change, only an overall shift in the eigenvalues of the Hamiltonian.
Condition \eqref{conda} is hence a simple alternative to a condition based on norms of $\widetilde H_\mathrm{eff}$, and occurs in the context of standard adiabatic elimination---the $\boldsymbol{\omega}$ of the interaction-picture Hamiltonian $H_\mathrm{I}$ in Eq.~\eqref{eq:HI} that gave the effective Hamiltonian of Eq.~\eqref{eq:Heff} from adiabatic elimination satisfies condition \eqref{conda}.
In the examples in the next section, we will explore the consequences of the different choices of $\widetilde{\omega}$ on the accuracy of the approximate solutions.

\section{Examples}\label{sec:example}
We now present several examples that illustrate our method: a three-level atom, a four-level atom, two atoms with Rydberg-blockade, and lastly, two three-level atoms.

\subsection{A three-level atom}\label{sec:EX1atom}

In order to check the validity of, and illustrate the differences between, the
adiabatic-elimination approximation and the first-order Markov approximation, 
we apply them to a three-level system in a $\Lambda$ configuration as depicted in
Fig.~\ref{fig:Lambda}. 
For a quantitative statement about the quality of the approximations, we compare the 
resulting Rabi frequency $\Omega_\mathrm{R}$ for population transfer between $|\mathrm{g}\rangle$ and $|\mathrm{t}\rangle$.
Here, the exact $\hbar\Omega_\mathrm{R}$ is the smallest spacing between the eigenvalues of $H_\mathrm{I}$, and is approximated by the difference of the two eigenvalues of the $\widetilde H_\mathrm{eff}$ under consideration.
We first consider the case of zero overall detuning, $\delta=0$.
Expanding the solutions under the various approximations to order 
$x^4$, where $x=\bigl(\Omega_0^2+\Omega_1^2\bigr)/4\Delta^2$,
gives the effective Rabi frequencies found in Table \ref{tab:1}(a).

Regardless of the choice of conditions \eqref{conda}-\eqref{condc}, the zeroth-order Markov approximation gives the correct Rabi frequency up to order $x$, while the first-order solutions give accuracy up to order $x^2$.
With a judicious choice of interaction picture, one can improve the approximations---for example, under the zeroth-order Markov approximation, employing condition \eqref{condc} in fact gives the \emph{exact} Rabi frequency.
However, we believe this is a coincidence rather than a general fact (see the $\delta\neq 0$ case discussed below).
A more reliable conclusion one can draw is that the first-order Markov approximation gives better accuracy, and that whether we choose condition \eqref{conda}, \eqref{condb}, or \eqref{condc} matters little.
In fact, one might even consider minimizing $\Vert\widetilde H_\mathrm{eff}\Vert$ for other norms (for example, the Hilbert-Schmidt norm).
The near-equivalence between the conditions is reassuring, since very often, condition \eqref{conda} is the easiest among the three possibilities to apply.
It also validates the common choice of condition \eqref{conda} in standard adiabatic elimination.

\setlength{\tabcolsep}{8pt}
\renewcommand{\arraystretch}{1.5}
\newcommand{\tabht}{0.5}
\begin{table*}
\centering
\caption{\label{tab:1}Rabi frequencies under various approximations for the example of a three-level atom with zero and nonzero overall detunings (see Sec.~\ref{sec:EX1atom}). Here, $x=$$\mbox{\footnotesize$\displaystyle\frac{\Omega_0^2+\Omega_1^2}{4\Delta^2}$}$ and
$\alpha=$$\mbox{\footnotesize$\displaystyle\frac{\Omega_0^2-\Omega_1^2}{\Omega_0^2+\Omega_1^2}$}\neq 0$.
}
\begin{tabular}{|r|r|l|}
\multicolumn{3}{@{}l}{(a) $\delta=0$}\\[1ex]
\hline
\multicolumn{2}{|l|}{exact solution}&$\frac{1}{\Delta}\Omega_\mathrm{R}=x-x^2+2x^3-5x^4+O(x^5)=\frac{1}{2}{\left(\sqrt{1+4x}-1\right)}$\\[\tabht ex]
\hline
\multirow{3}{*}{0th-order Markov}&Condition \eqref{conda}&$\frac{1}{\Delta} \Omega_\mathrm{R}= x$\\[\tabht ex]
\cline{2-3}
&Condition \eqref{condb}&$\frac{1}{\Delta} \Omega_\mathrm{R}=x-\frac{1}{2}x^2+\frac{1}{2}x^3-\frac{5}{8}x^4+O(x^5)=\sqrt{1+2x}-1$\\[\tabht ex]
\cline{2-3}
&Condition \eqref{condc}&$\frac{1}{\Delta} \Omega_\mathrm{R}=x-x^2+2x^3-5x^4+O(x^5)=\frac{1}{2}{\left(\sqrt{1+4x}-1\right)}$\\[\tabht ex]
\hline
\multirow{3}{*}{1st-order Markov}&Condition \eqref{conda}& $\frac{1}{\Delta} \Omega_\mathrm{R} =  x-x^2+x^3-x^4+O(x^5)=\frac{x}{1+x}$\\[\tabht ex]
\cline{2-3}
&Condition \eqref{condb}& $\frac{1}{\Delta}\Omega_\mathrm{R} =   x-x^2+\frac{7}{4}x^3-\frac{15}{4}x^4+O(x^5)$\\[\tabht ex]
\cline{2-3}
&Condition \eqref{condc}&$\frac{1}{\Delta}\Omega_\mathrm{R}=x-x^2+x^3-x^4+O(x^5)=\frac{x}{1+x}$\\[\tabht ex]
\hline
\multicolumn{3}{l}{ }\\
\multicolumn{3}{@{}l}{(b) $\mbox{\footnotesize$\displaystyle\delta=\frac{\Omega_1^2-\Omega_0^2}{4\Delta}$}$}\\[1.5ex]\hline
\multicolumn{2}{|l|}{exact solution}&$\frac{1}{\Delta}\Omega_\mathrm{R}=\sqrt{1-\alpha^2}{\left\{x-x^2+{\left[2+O(\alpha^2)\right]}x^3+O(x^4)\right\}}$\\[\tabht ex]
\hline
\multirow{3}{*}{0th-order Markov}&Condition \eqref{conda}& $\frac{1}{\Delta} \Omega_\mathrm{R}=\sqrt{1-\alpha^2} \,x$\\[\tabht ex]
\cline{2-3}
&Condition \eqref{condb}& $\frac{1}{\Delta} \Omega_\mathrm{R}=\sqrt{1-\alpha^2}{\left\{x-\frac{1}{2}x^2+{\left[\frac{1}{2}+O(\alpha^2)\right]}x^3+O(x^4)\right\}}$\\[\tabht ex]
\cline{2-3}
&Condition \eqref{condc}& $\frac{1}{\Delta} \Omega_\mathrm{R}=\sqrt{1-\alpha^2}{\left\{x-{\left[1+O(\alpha^2)\right]}x^2+{\left[2+O(\alpha^2)\right]}x^3+O(x^4)\right\}}$\\[\tabht ex]
\hline
\end{tabular}
\end{table*}

For comparison, we consider a case where $\delta\neq 0$. In many experiments that aim at full population transfer between the ground state and the target state, $\delta$ is adjusted to compensate for the light shifts of the atomic levels, due to the presence of the lasers, to give a resonant two-photon transition. 
Often, the choice is $\delta=(\Omega_1^2-\Omega_0^2)/4\Delta$, a value designed to make the diagonal entries of the effective Hamiltonian of Eq.~\eqref{eq:Heff} from adiabatic elimination equal. 
This ensures a resonant process, to the accuracy justified by the adiabatic-elimination approximation. 
For this value of $\delta$, the predicted Rabi frequencies for the exact solution and for the zeroth-order Markov approximation under the different conditions are listed in Table \ref{tab:1}(b).

As in the $\delta=0$ example above, the zeroth-order approximation gives the correct answer to linear order in $x$, regardless of the choice of condition.
That condition \eqref{condc} gives the exact Rabi frequency is no longer true here, although for small $\alpha$, it still provides the most accurate prediction.
One can also verify that the first-order Markov approximation again gives the correct value for the Rabi frequency up to second order in $x$.
A pertinent remark is that, in this case where $\delta\neq 0$, conditions \eqref{condb} and \eqref{condc} are rather more complicated to impose than condition \eqref{conda}. 
This lends justification to improving the description of the system by going to higher-order Markov approximations using only the simplest condition \eqref{conda}, instead of correcting the zeroth-order approximation by a more complicated choice of $\widetilde\omega$. 
We will see this same conclusion in the other examples discussed below.

\begin{figure}
\centering
\includegraphics[width=\columnwidth]{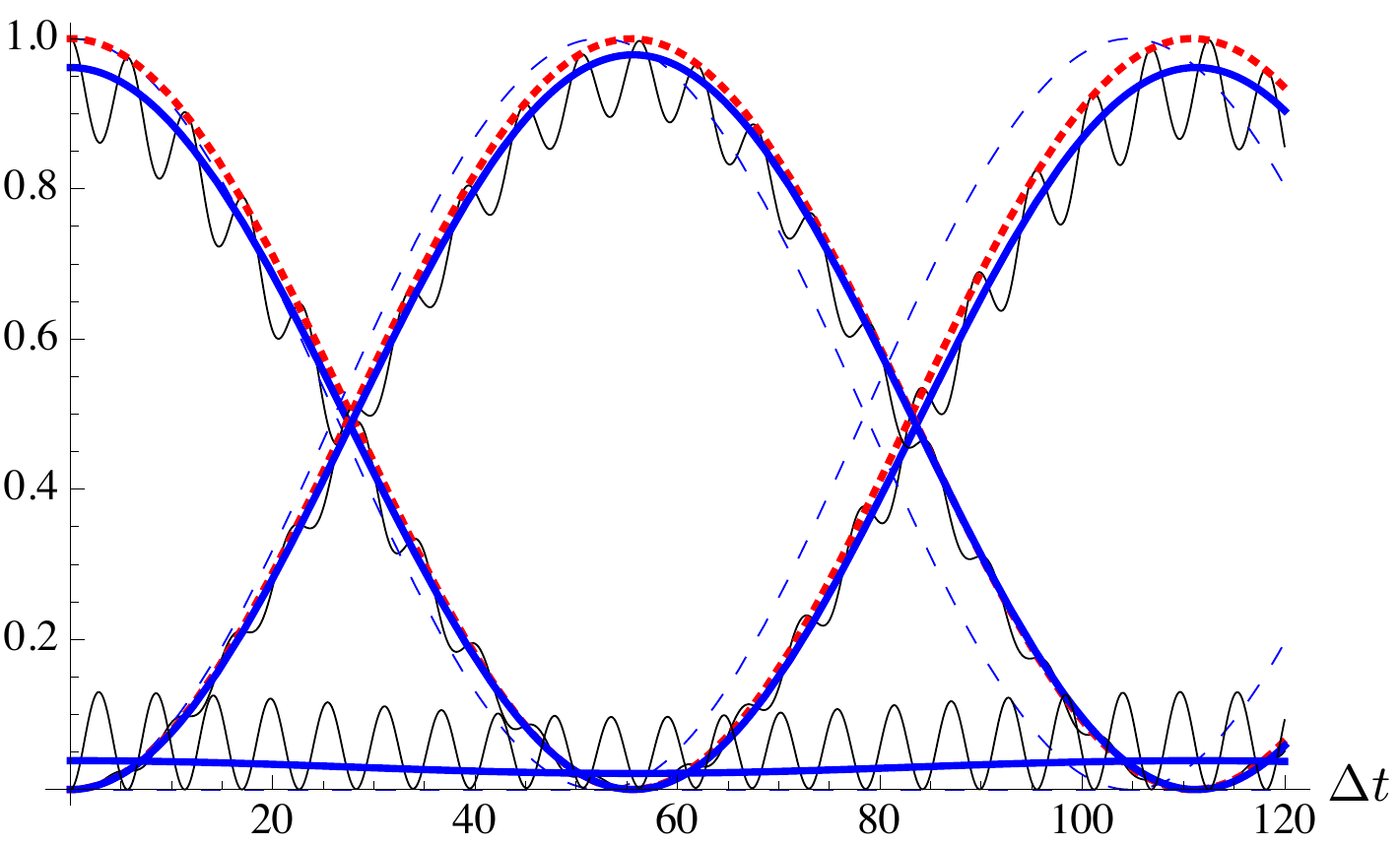}
\caption{\label{fig:3Levcomp}
  Population probabilities for a two-photon Raman transition in a three-level atom (see Sec.~\ref{sec:EX1atom}) as a function of $\Delta t$, for Rabi frequencies $\Omega_0=0.4\Delta$ and $\Omega_1=0.3 \Delta$, 
  and overall detuning $\delta=(\Omega_1^2-\Omega_0^2)/4\Delta$.
  The curves that start close to $1$ show the population of the ground state $|\mathrm{g}\rangle$;
  those that start at $0$ and increase to about $1$ before decreasing
  again are for the target state $|\mathrm{t}\rangle$; those that start near
  $0$ and never grow to large values are for the intermediate
  state $|\mathrm{e}\rangle$. 
  The solid black curves with wiggles plot the exact solution; 
  the thin dashed blue curves show the zeroth-order 
  Markov approximation with condition \eqref{conda} (standard adiabatic elimination);
  the dotted red curves are also for the zeroth-order Markov approximation, but with condition \eqref{condc};
  the solid thick blue curves show the first-order Markov approximation with condition \eqref{conda}.
  }
\end{figure}

To further illustrate matters, we examine the predictions for the evolution of the populations in the three states under the different approximations for the case with Rabi frequencies $\Omega_0=0.4\Delta$ and $\Omega_1=0.3 \Delta$,
and overall detuning $\delta=(\Omega_1^2-\Omega_0^2)/4\Delta$.
The results are plotted in Fig.~\ref{fig:3Levcomp}.
Standard adiabatic elimination (thin dashed blue lines in the plot), equivalent to the zeroth-order Markov approximation with condition \eqref{conda}, gives  a period for the Rabi oscillation that is about
6\% short of the exact value.
The zeroth-order Markov solution with condition \eqref{condc}, on the other hand, gives the exact Rabi frequency, but makes no statement about the population of the intermediate state.
In contrast, the first-order Markov approximation, even with the simplest condition \eqref{conda}, performs very well, and provides a reasonable estimate of the intermediate state population.

Note that all the approximations show the effect of the coarse graining:
They do not reproduce the high-frequency modulation of the exact
solution. 
This coarse graining is also behind the fact that the population of the ground state, under the first-order Markov approximation, does not start at 1 at time $t=0$. 
Instead, it begins with a value that can be thought of as the average over the initial coarse-grained time step, and is consistent with having an initial nonzero population in the intermediate state in compliance with Eq.~\eqref{eq:epseps}.

\subsection{A four-level atom}\label{sec:EX3photons}

\begin{figure*}
\centering
\includegraphics[scale=0.60]{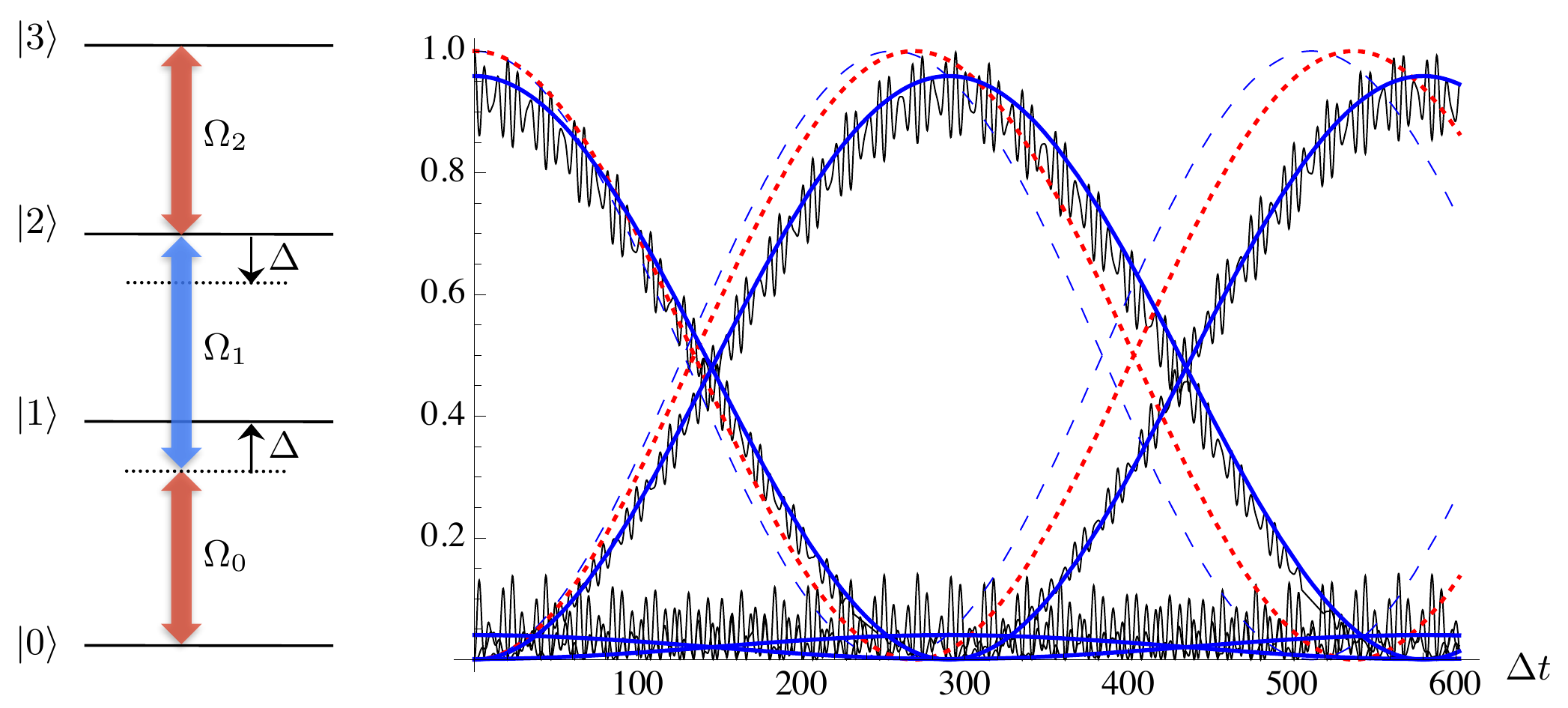}
\caption{\label{fig:3photonTr}
  Population probabilities for a three-photon Raman transition in a four-level
  atom (see Sec.~\ref{sec:EX3photons}), as a function of $ \Delta t$. 
  The plots are for $\Omega_0=\Omega_2=0.4\Delta$ and $\Omega_1=0.3\Delta$, with two relevant and two
  auxiliary levels in the system. 
  The curves that start near $1$ show the probability of the ground state
  $\ket{0}$;
  the curves that start at $0$ and increase to about $1$ before decreasing
  again show the probability of the target state $\ket{3}$; 
  and the curves that start near $0$ and never grow to large values show the
  probabilities of the other two intermediate states.
The solid black curves  with wiggles show the full solution, 
  the dashed blue curves are for the zeroth-order Markov approximation with condition \eqref{conda}, the dotted red curves are for the zeroth-order Markov approximation with condition \eqref{condc}
  and the solid thick blue curves are for the first-order Markov approximation with condition \eqref{conda}.}
\end{figure*}

For the single three-level atom in Section \ref{sec:EX1atom}, one hardly needs
an approximate treatment. Nevertheless, the reduction of the description to the relevant
two levels is a helpful simplification and the coarse-grained probabilities
thus obtained are often all one needs to know to determine the important
parameter range for an experiment. 
When more levels are involved, a full treatment may no longer be possible or
contain too many irrelevant details.

As an example of such a more complex situation, we first consider a four-level atom where the the transition between the ground state $\ket{0}$ and target state $\ket{3}$ is achieved with a three-photon process via two far-detuned intermediate states $\ket{1}$ and $\ket{2}$; see the level scheme in Fig.~\ref{fig:3photonTr}. The interaction-picture Hamiltonian of Eq.~\eqref{eq:HINlevel} is
\begin{equation}\label{eq:3phHint1}
H_\mathrm{I} =
\hbar{\left( \begin{array}{cccc} 
    0 & \frac{1}{2}\Omega_0 & 0 & 0 \\[0.5ex]
    \frac{1}{2}\Omega_0 & \Delta_0 & \frac{1}{2}\Omega_1 & 0 \\
    0 & \frac{1}{2}\Omega_1& \Delta_0+\Delta_1 & \frac{1}{2}\Omega_2 \\[0.5ex]
    0 & 0 &  \frac{1}{2}\Omega_2 & \Delta_0+\Delta_1+\Delta_2  
  \end{array} \right)}.
\end{equation}
When the overall detuning $|\Delta_0+\Delta_1+\Delta_2|$ is small, the relevant states are $\ket{0}$ and $\ket{3}$. 
The light shifts of states $\ket{0}$ and $\ket{3}$ are $\Omega_0^2/4\Delta_0$ and $-\Omega_2^2/4\Delta_2$, respectively. 
For simplicity, we let $\Delta_0=-\Delta_2=\Delta$ and $\Omega_0=\Omega_2$ so that the two relevant states have the same magnitude for the light shift. We also set $\Delta_1=0$ for a resonant three-photon transition. After re-organizing the Hamiltonian into relevant and irrelevant sectors, we obtain
\begin{equation}\label{eq:3phHint2}
H_\mathrm{I} =
\hbar{\left( \begin{array}{cc|cc} 
    0 & 0 & \frac{1}{2}\Omega_0 & 0 \\[0.5ex]
   0 & 0 & 0 & \frac{1}{2}\Omega_2 \\
    [1ex] \hline\rule{0pt}{3ex}
     \frac{1}{2}\Omega_0 & 0& \Delta & \frac{1}{2}\Omega_1 \\[0.5ex]
    0 & \frac{1}{2}\Omega_2 &  \frac{1}{2}\Omega_1 & \Delta  
  \end{array} \right)},
\end{equation}
now written in the basis $\{|0\rangle,|3\rangle,|1\rangle,|2\rangle\}$.

The zeroth-order (adiabatic elimination) and first-order Markov approximations can be applied to this system with the $4=2+2$ split of the Hamiltonian in Eq.~\eqref{eq:3phHint2}. An example of the population of states using different approximations is given in Fig.~\ref{fig:3photonTr}. It shows that the adiabatic elimination with condition \eqref{condc} works better than condition \eqref{conda}. However, by going to the first-order Markov approximation with condition \eqref{conda}, the accuracy of the approximation improves significantly and it gives the average of the populations in the relevant states almost exactly. 
This conclusion holds even for the general Hamiltonian in Eq.~\eqref{eq:3phHint1} without the simplification of setting $\Delta_0 = -\Delta_2$ or having equality between $\Omega_0$ and $\Omega_2$.

\subsection{Two atoms with Rydberg blockade}\label{sec:EX2atomsRyd}

\begin{figure*}
\centering
\includegraphics[scale=0.55]{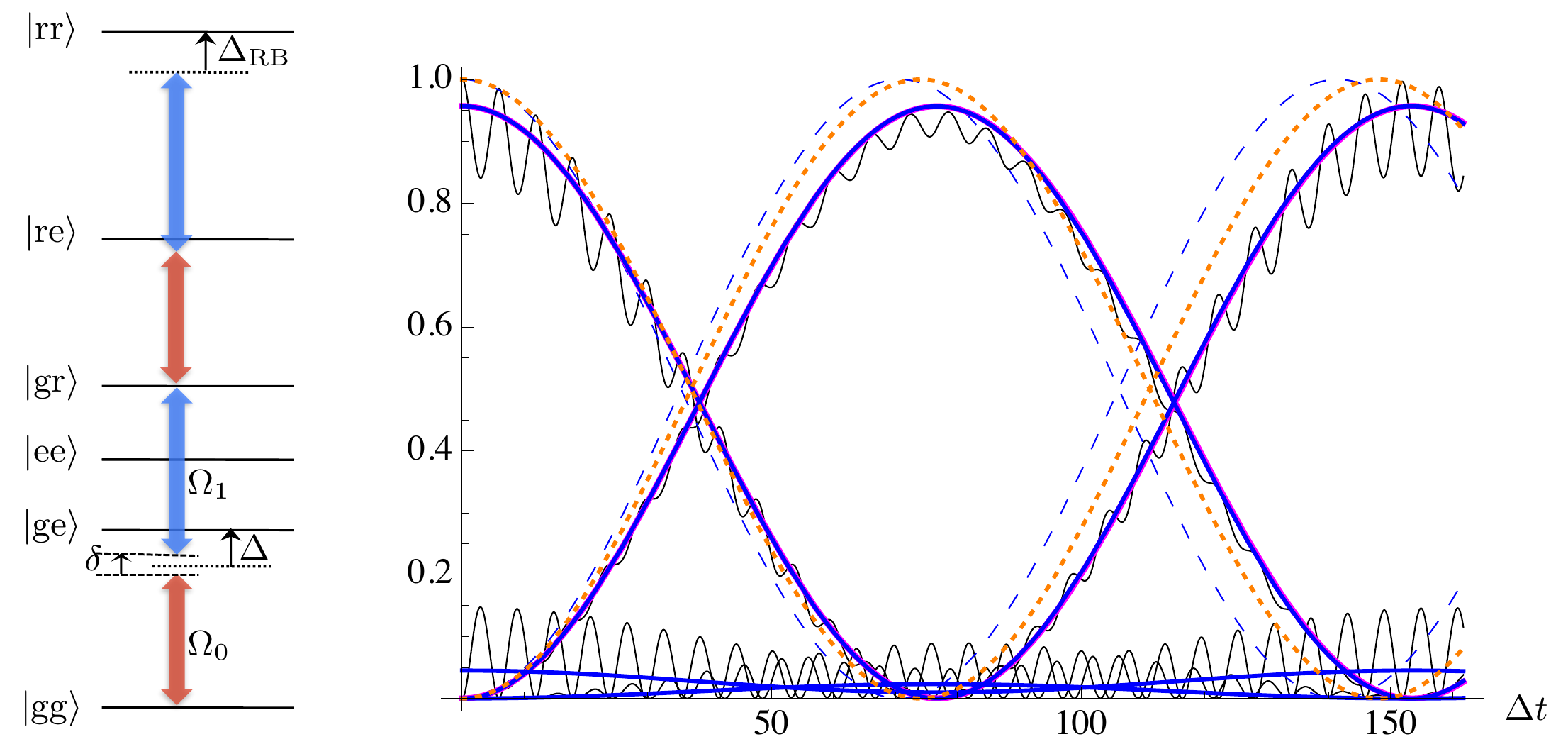}
\caption{\label{fig:2AtomRB}
  Population probabilities for a three-level cascade Raman transition in a two-atom system with Rydberg blockade (see Sec.~\ref{sec:EX2atomsRyd}), as a function of $ \Delta t$. 
  The plots are for $\Omega_0=0.3\Delta$, $\Omega_1=0.2\Delta$, 
  $\delta=(\Omega_1^2-\Omega_0^2)/4\Delta$ and $\Delta_{\mathrm{RB}}=5\Delta$.  
  The curves that start near $1$ show the probability of the double ground state
  $|\mathrm{g}\mathrm{g}\rangle$;
  the curves that start at $0$ and increase to about $1$ before decreasing
  again show the probability of the target state $|\mathrm{g}\mathrm{r}\rangle$; 
  and the curves that start near $0$ and never grow to large values show the
  probabilities of the other states.
The solid black curves with wiggles show the full solution; 
  the dashed blue curves are for the zeroth-order Markov approximation with condition \eqref{conda} and the $6=2+4$ split; the dotted orange curves are Rabi oscillations drawn with Rabi frequency $\sqrt{2}\Omega_0\Omega_1/(2\Delta)$ (which is the often-used expression for the effective Rabi frequency with $\sqrt{N}$ enhancement for a $N$-atom Rydberg excitation \cite{Heidemann});
  and the solid thick blue and purple curves are for the first-order Markov approximation using condition
\eqref{conda} with a $6=2+2+2$ split and a $6=2+4$ split, respectively. They are hardly distinguishable from each other in this figure.}
\end{figure*}

Here, we consider two identical three-level atoms, each in a cascade configuration, where the highest energy level for each atom is a Rydberg state---a state with a large principal quantum number and thus
large electric dipole moment. The atoms are so close
to each other that a Rydberg blockade \cite{Gallagher} happens; see the level scheme in Fig.~\ref{fig:2AtomRB}.

Initially both atoms are in the ground state $|\mathrm{g}\rangle$ and we have 
$|\mathrm{g}\mathrm{g}\rangle=|\mathrm{g}\rangle\otimes|\mathrm{g}\rangle$ for the initial state of the two-atom
system.  
With the two driving lasers coupling in the same way and with the same
strength to both atoms, the two-atom state will be invariant under the
permutation of the atoms for all later times.
We use a shorthand notation in which, for instance, $|\mathrm{g}\mathrm{r}\rangle$ denotes the
state with one atom in the ground state and the other atom in the Rydberg
state,  
$|\mathrm{g}\mathrm{r}\rangle=(|\mathrm{g}\rangle\otimes|\mathrm{r}\rangle+|\mathrm{r}\rangle\otimes|\mathrm{g}\rangle)/\sqrt{2}$.
In total, then, six two-atom states participate in the evolution:
$|\mathrm{g}\mathrm{g} \rangle$, $|\mathrm{g}\mathrm{r}\rangle$, $|\mathrm{g}\mathrm{e}\rangle$, $|\mathrm{r}\mathrm{e}\rangle$,
$|\mathrm{ee} \rangle$, and $|\mathrm{r}\mathrm{r}\rangle$. 
Referring to this order, the interaction-picture Hamiltonian of Eq.~\eqref{eq:HINlevel} is
\begin{equation}\label{eq:2atHint}
H_\mathrm{I} =
\hbar{\left( \begin{array}{@{}cc|cccc@{}} 
    0 & 0 & \frac{\Omega_0}{\sqrt2} & 0 & 0 & 0 \\[0.5ex]
    0 & \delta & \frac{\Omega_1}{2} 
        & \frac{\Omega_0}{2} & 0 & 0 \\[1ex] \hline\rule{0pt}{3ex}
    \frac{\Omega_0}{\sqrt2} & \frac{\Omega_1}{2} &
          \Delta + \frac{\delta}{2} & 0 & 
      \frac{\Omega_0}{\sqrt2} & 0\\[0.5ex]
    0 & \frac{\Omega_0}{2} & 0 & \Delta +\frac32 \delta &
    \frac{\Omega_1}{\sqrt2} & \frac{\Omega_1}{\sqrt2}  \\[1ex] 
     0 & 0 & \frac{\Omega_0}{\sqrt2} & \frac{\Omega_1}{\sqrt2} & 
     2 \Delta + \delta & 0 \\[0.5ex]
     0 & 0 & 0 & \frac{\Omega_1}{\sqrt2} & 0 &
       \Delta_\mathrm{RB} + 2 \delta
  \end{array} \right)},
\end{equation}
where, as before, $\Delta=(\Delta_0+\Delta_1)/2$ is the average detuning for a single three-level atom, and $\delta=\Delta_0-\Delta_1$ is the overall detuning.
The energy of the double Rydberg state $|\mathrm{r}\mathrm{r}\rangle $ is shifted by an energy 
$\hbar\Delta{_\mathrm{RB}}$ due to the strong dipole-dipole interaction between Rydberg levels.
When $\vert\Delta_\mathrm{RB}\vert\gg\Omega_0,\Omega_1$, the Rydberg blockade mechanism prevents both atoms from being excited to the Rydberg
state at the same time, and this mechanism can be used to entangle atoms and
implement quantum gates \cite{Jaksch,Brion2}. 
As a typical value of this energy shift, we take
$\Delta_\mathrm{RB}$ to be a few times the size of $\Delta$.
Since the double Rydberg state $|\mathrm{r}\mathrm{r}\rangle$ is not going to be populated, the
goal is a population transfer from the double ground state $|\mathrm{g}\mathrm{g}\rangle $ to
the one-Rydberg-atom target state $|\mathrm{g}\mathrm{r}\rangle$, i.e., 
a transfer between the two states to
which the two left columns and the two top rows in Eq.~\eqref{eq:2atHint} refer.

There are two pairs of resonant transitions: 
$|\mathrm{g}\mathrm{g}\rangle \leftrightarrow |\mathrm{g}\mathrm{r} \rangle$ and 
$|\mathrm{g}\mathrm{e}\rangle \leftrightarrow |\mathrm{r}\mathrm{e} \rangle$.
However, if the initial state is $\ket{\mathrm{g}\mathrm{g}}$, there are only two relevant states $\ket{\mathrm{g}\mathrm{g}}$ and $\ket{\mathrm{g}\mathrm{r}}$, because states $|\mathrm{g}\mathrm{e}\rangle$ and $|\mathrm{r}\mathrm{e}\rangle$ are barely populated although they are resonantly coupled to each other.
Correspondingly, we have a ${6=2+4}$ split of the Hamiltonian into relevant and irrelevant sectors, as
indicated in Eq.~\eqref{eq:2atHint}. The major computational complexity is given by inverting the $4\times4$ matrix $\boldsymbol{\Delta}$ for the irrelevant sector, as required by Eq.~\eqref{eq:HeffM}. 

Alternatively, it is also possible to apply a two-step Markov approximation, i.e., first
eliminating the last two states to obtain a four-dimensional effective Hamiltonian, and then approximate the system with a 4=2+2 split for an effective description of the two relevant states. We refer to this as a $6=2+2+2$ split. 
The choice of interaction picture is the same as in the case of the 6=2+4 split, since there are only two relevant states even if only two of the irrelevant states are eliminated in the first step. 
The computational complexity is much reduced with this two-step approximation, because at each step one only needs to invert $2\times2$ matrices rather than $4\times4$ matrices. 
But this has no additional benefit if both steps are done with the zeroth-order Markov approximation; it gives the same result as applying the zeroth-order Markov approximation directly to the ${6=2+4}$ split. However, a significant improvement can be obtained using the zeroth-order Markov approximation in the first step and the first-order Markov approximation in the second step. 
Using the first-order Markov approximation in the first step would not improve the approximation significantly as the last two states $\ket{\mathrm{ee}}$ and $\ket{\mathrm{r}\mathrm{r}}$ are much farther detuned than states $\ket{\mathrm{g}\mathrm{e}}$ and $\ket{\mathrm{r}\mathrm{e}}$. 

The plots in Fig.~\ref{fig:2AtomRB} show the differences between the zeroth-order Markov approximation (adiabatic elimination) with condition \eqref{conda} and condition \eqref{condc} for the $6=2+4$ split, where the approximation with condition \eqref{condc} works better than the approximation with condition \eqref{conda}. The first-order Markov approximation with the $6=2+2+2$ split works as well as the first-order Markov approximation with the $6=2+4$ split, and both of them describe the overall oscillation of the two relevant states very well. However, the effective Hamiltonian with the $6=2+2+2$ split is much easier to obtain than with the $6=2+4$ split.
Since the first-order Markov approximation is insensitive to the choice of interaction picture, we apply the simplest condition \eqref{conda} for the approximate solution using either split.

\subsection{Two three-level atoms}\label{sec:EX2atoms}
\begin{figure}
\centering
\includegraphics[scale=0.62]{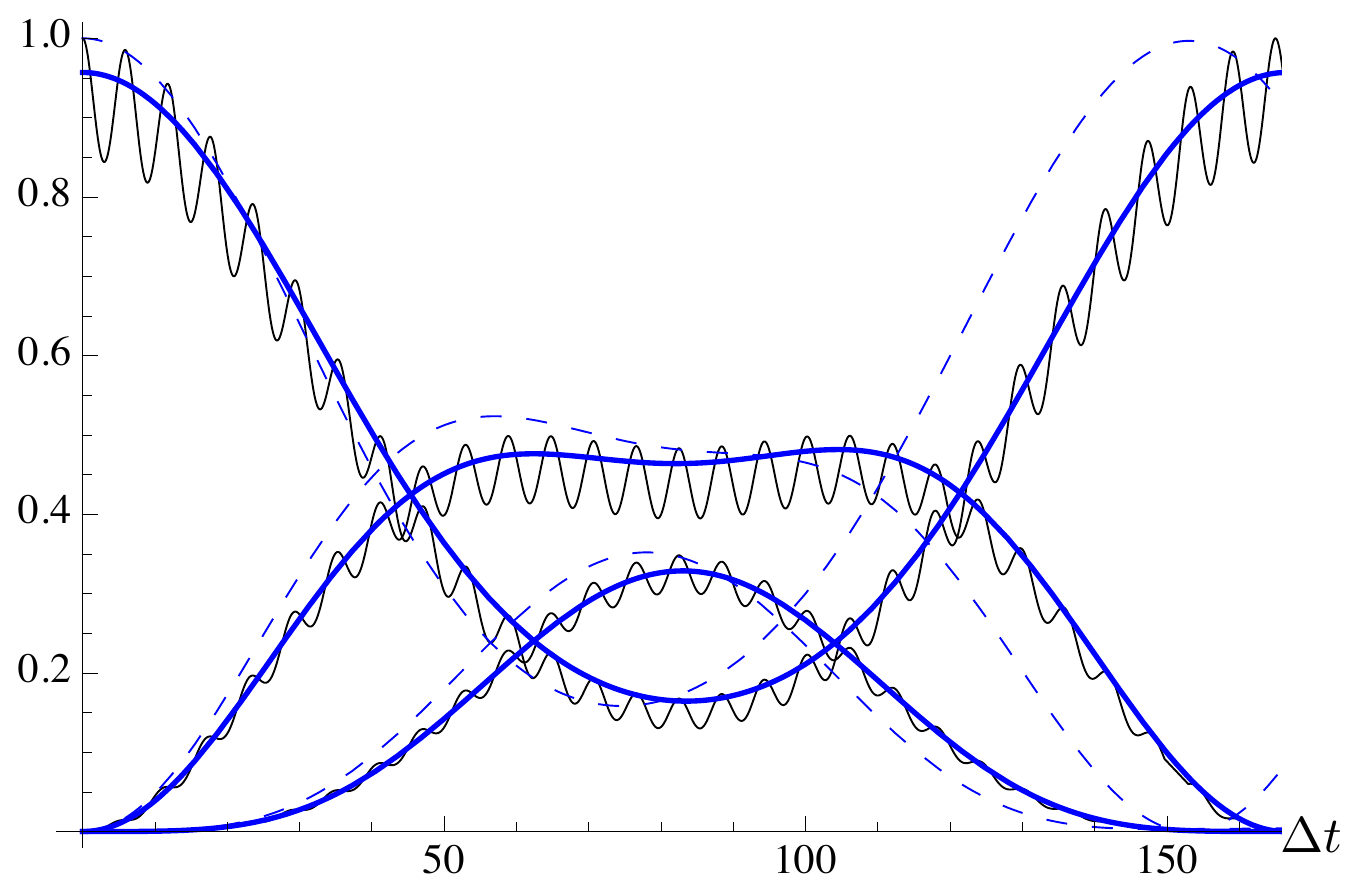}
\caption{\label{fig:2AtomNoRB}
  Population probabilities of the relevant states for a 
  three-level cascade Raman transition in a two-atom system without Rydberg blockade (see Sec.~\ref{sec:EX2atoms}), as a function of $ \Delta t$. 
  The plots are for $\Omega_0=0.3\Delta$, $\Omega_1=0.2\Delta$ and 
  $\delta=(\Omega_1^2-\Omega_0^2)/4\Delta$.
  The curves that start near $1$ show the probability of the double ground state
  $|00\rangle$;
  the curves that start at $0$ with a single peak show the probability of the two-atom target state $|\mathrm{tt}\rangle$; 
  and the curves that start near $0$ with double peaks show the
  probabilities of the one-atom target state $\ket{\mathrm{gt}}$.
The solid black curves  with wiggles show the full solution; the dashed blue curves are the solutions when the three irrelevant states are adiabatically eliminated with condition \eqref{conda};
  and the solid thick blue curves are for the first-order Markov approximation with condition \eqref{conda}.}
\end{figure}

Here, we consider two identical three-level atoms in either the $\Lambda$ or the cascade configuration, where there is no blockade of the doubly excited states. Thus, we can define the states in the same way as in Section \ref{sec:EX2atomsRyd} with no Rydberg blockade for the target state $\ket{\mathrm{tt}}$, i.e. $\Delta_\mathrm{RB}=0$. We re-arrange the basis in the order: $|\mathrm{gg} \rangle$, $|\mathrm{gt}\rangle$, $\ket{\mathrm{tt}}$, $|\mathrm{g}\mathrm{e}\rangle$, $|\mathrm{t}\mathrm{e}\rangle$ and $|\mathrm{ee} \rangle$, and the interaction-picture Hamiltonian is
\begin{equation}\label{eq:2atHint-2}
H_\mathrm{I} =
\hbar{\left( \begin{array}{ccc|ccc} 
    0 & 0 & 0 & \frac{\Omega_0}{\sqrt2} & 0 & 0 \\[0.5ex]
    0 & \delta & 0 
       & \frac{\Omega_1}{2} & \frac{\Omega_0}{2} & 0 \\
        0 & 0 & 2\delta & 0 & \frac{\Omega_1}{\sqrt{2}} & 0\\[1ex] \hline\rule{0pt}{3ex}
    \frac{\Omega_0}{\sqrt2} & \frac{\Omega_1}{2} & 0 &
          \Delta + \frac{\delta}{2} & 0 & 
      \frac{\Omega_0}{\sqrt2} \\[0.5ex]
    0 & \frac{\Omega_0}{2}  & \frac{\Omega_1}{\sqrt2} & 0 & \Delta +\frac32 \delta &
    \frac{\Omega_1}{\sqrt2}  \\[1ex] 
     0 & 0 & 0 & \frac{\Omega_0}{\sqrt2} & \frac{\Omega_1}{\sqrt2} & 
     2 \Delta + \delta \\[0.5ex]
  \end{array} \right)},
\end{equation}
where a $6=3+3$ split is shown and the three relevant states are $|\mathrm{gg} \rangle$, $|\mathrm{gt}\rangle$ and $\ket{\mathrm{tt}}$. The plots in Fig.~\ref{fig:2AtomNoRB} show the difference between adiabatic elimination with condition (\ref{eq:optcond}a) and the first-order Markov approximation with condition (\ref{eq:optcond}a). Here, too, we observe that the first-order Markov approximation gives very reliable results.

\section{Summary and outlook}\label{sec:outlook}

We have described a procedure to derive a systematic hierarchy of approximations for effective descriptions of multi-level, multi-photon Raman processes.
The lowest-order approximation is the well-known adiabatic elimination method, and we managed to give answers to all four of the questions asked in Sec.~\ref{sec:problems} regarding issues for adiabatic elimination.
The first-order Markov approximation provides a notable improvement to adiabatic elimination.
If used with the simplest condition \eqref{conda} for the choice of interaction picture, it gives an accurate solution even for rather complicated systems while not increasing the implementation complexity by much.

This is not the end of the story, however.
It will be instructive to apply the first-order Markov approximation to other
more complex situations where standard adiabatic elimination is currently the
method of choice. 
One should also extend the method to include spontaneous emission or, more
generally, investigate how the method can be modified for open quantum systems.
Lastly, we mention that a systematic approximation without the use of adiabatic elimination is possible as well \cite{Raman-nonadiabatic}.

\begin{acknowledgments}
This work is supported by the National Research Foundation and the Ministry of
Education, Singapore. 
V.~P. would like to thank the Studienstiftung des deutschen Volkes for financial support
and the Centre for Quantum Technologies for the hospitality experienced during an internship.
\end{acknowledgments}



%

\end{document}